# The power-law reaction rate coefficient for an elementary bimolecular reaction

Yin Cangtao and Du Jiulin[*]

*Department of Physics, School of Science, Tianjin University, Tianjin 300072, China*

**Abstract:** The power-law TST reaction rate coefficient for an elementary bimolecular reaction is studied when the reaction takes place in a nonequilibrium system with power-law distributions. We derive a generalized TST rate coefficient, which not only depends on a power-law parameter but also on the reaction coordinate frequency of transition state. The numerical analyses show a very strong dependence of the TST rate coefficient on the power-law parameter, and clearly indicate that a tiny deviation from unity in the parameter (thus from a Boltzmann-Gibbs distribution) would result in significant changes in the rate coefficient. We take an elementary reaction, $F+H_2 \to FH+H$, as an application example to calculate the reaction rate coefficient, and yield the rate values being exactly agreement with the measurement values in all the experimental studies in temperature range 190~765K.

## 1. Introduction

The calculation of reaction rate coefficient is an interdiscipline of nonlinear science, and it is very important for studying and understanding many basic problems in many different physical, chemical, biological and technical processes. There are various reaction rate theories that have been developed to calculate the reaction rate coefficient, among which transition state theory (TST) is the most basic one. TST has made it possible to obtain quick estimates for the reaction rates of a broad variety of processes in natural science and thus became a cornerstone and a core of the reaction rate theory. For an elementary reaction process [1], for example,

$$A + BC \to AB+C, \qquad (1)$$

TST states that the reactants form firstly the activated complex $A \cdots B \cdots C$ (i.e. the transition state) and then become the products, where the old bond B-C is stretching and the new bond A-B is forming. In typical applications, the regions of reactant and product are associated with minima of the potential energy and are separated by a high barrier where the potential energy has a saddle point. In the neighborhood of the saddle point, the potential energy has a maximum along the reaction coordinate [2]. It has been assumed that after the reactant molecules colliding with each other cross over the transition state, they no longer return and definitely become the product. Respectively, the reactant molecules and the transition state are assumed to comply with a statistical distribution at a thermodynamic equilibrium state, and thus a Boltzmann-Gibbs (BG) distribution is considered to be the statistical base of TST. In this way, TST reaction rate coefficient is conventionally written by a form with the exponential law [3],

$$k_{TST} = \frac{k_B T}{h} Z^{-1} \exp(-\beta \Delta \varepsilon_0), \quad (2)$$

where $Z$ is a partition function, $\Delta \varepsilon_0$ is difference of basic energies between transition state and reactants, $h$ is Planck constant, $k_B$ is Boltzman constant, $T$ is temperature and $\beta = (k_B T)^{-1}$ is Lagrangian multiplier.

---

[*] Corresponding author. E-mail address: 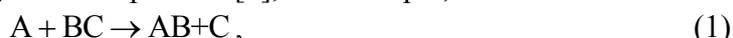jldu@tju.edu.cn

However, the statistical property of complex systems away from equilibrium does not always follow BG statistics and therefore does not have an exponential-law distribution. A lot of theoretical work and experimental studies on physical, chemical, biological and technical processes taking place in complex systems have shown that the statistical property often follows power-law distributions (see [4] and the references therein). Here we can write some of the distributions such as the form known in nonextensive statistical mechanics [5]. We can introduce the power-law $\nu$-distribution,

$$P(\varepsilon) \sim [1-(\nu-1)\beta\varepsilon]^{1/(\nu-1)}, \quad (3)$$

if the variable $\varepsilon$ is small. Or we can write $P(\varepsilon) \sim \varepsilon^{-\alpha}$ if the variable $\varepsilon$ is large [4]. This power-law $\nu$-distribution represents the statistical property of a system being at nonequilibrium stationary-state [6, 7]. Eq.(3) is reduced to a BG distribution if the $\nu$-parameter is set $\nu \to 1$, where the parameter $\nu \neq 1$ measures a distance away from the equilibrium and it is a function of the energy [4]. The power-law distributions in complex systems have been found and noted prevalently in the processes such as single-molecule conformational dynamics [8, 9], reaction-diffusion processes [10], chemical reactions [11], combustion processes [12], gene expressions [13], cell reproductions [14], complex cellular networks [15], and small organic molecules [16] etc. In these processes, the reaction rate coefficients may be energy-dependent (and/or time-dependent [17, 18]) power-law forms [19, 20], which are beyond the scope of conventional TST reaction rate formulae with a BG exponential law. In these cases, the traditional rate formulae of TST reaction rate theory become invalid and so need to be modified. Recently, the conditions resulting power-law distributions were obtained by means of the stochastic dynamics on Brownian motion in complex systems, and a generalized Klein-Kramers equation and a generalized Smoluchowski equation for the systems with power-law distributions were developed [4]. Simultaneously, A TST reaction rate theory for the nonequilibrium systems with power-law distributions was also studied by using basic dynamical and statistical theory, and the generalized rate formulae were presented [19]. As we can imagine, it will be a complicated and exciting field in exploring the understanding of open nonequilibrium reaction rate theory.

The purpose of this work is to generalize a conventional TST reaction rate formula to a nonequilibrium system with the power-law $\nu$-distribution. In Sec.2, we study the power-law TST reaction rate coefficient of elementary reaction (1). In Sec.3, we make numerical analyses to show dependence of the power-law TST reaction rate coefficient on the quantities such as ν-parameter, temperature and reaction coordinate frequency etc. As an application example of the new TST formula, in Sec.4 we calculate the power-law reaction rate coefficients of F+H$_2$→HF+H reaction, compare them with the values in experiment studies, and determine the ν-parameter. Finally, in Sec.5 we give the conclusion and discussion.

## 2. The power-law TST rate coefficient for an elementary reaction

Usually, conventional TST reaction rate formulae are derived by means of using elementary reactions [1, 21]. As the first step of the generalization of TST rate formulae to systems with power-law distributions, we will follow the standard line of textbooks to derive the power-law TST reaction rate formula. Let us consider an elementary reaction, Eq.(1). The reaction process [1] carries out in the following mode,

$$\text{A+BC} \rightleftharpoons \text{X}^{\neq}(\text{A}\cdots\text{B}\cdots\text{C}) \rightarrow \text{AB+C}. \qquad (4)$$

In these reaction process, reactants firstly form activated complex $\text{A}\cdots\text{B}\cdots\text{C}$, i.e. transition state, and then become products, where the old bond between B-C is stretching and the new one in A-B is forming. Strictly speaking, equilibrium constants of the reaction are based on activities, not concentrations (or fugacities rather than pressures). An equilibrium constant expressed in terms of activities (or fugacities) is called a thermodynamic equilibrium constant. In elementary applications, the activities that occur in the equations are often approximately replaced by concentrations or pressures [22]. As usual, if $\text{X}^{\neq}$ denotes the transition state, TST holds that there is an equilibrium constant $K^{\neq}$ [1], given by the form

$$K^{\neq} = \frac{c^{\neq}}{c_A\, c_{BC}}, \quad \text{or} \quad c^{\neq} = K^{\neq} c_A\, c_{BC}, \qquad (5)$$

where $c^{\neq}$, $c_A$ and $c_{BC}$ are, respectively, concentration of activated complex $\text{X}^{\neq}$, reactant molecule A and reactant molecule BC. One can directly study this elementary reaction, without need to consider the details of state-state reactions. In such a mode, the reaction coordinate motion in transition state (i.e. the old bond between B-C is stretched and the new one in A-B is formed) can be separated from other motions such as translational, rotational and vibrational motion etc. The rate of overall reaction processes can be expressed as the rate of the reaction coordinate motion. Let $\omega_f$ be the reaction coordinate frequency (i.e. decomposition frequency of transition state), then $1/\omega_f$ is mean lifetime of transition state. The reaction rate $\upsilon_f$ is product of the concentration and the reaction coordinate frequency, i.e. $\upsilon_f = c^{\neq}\omega_f$. Using Eq. (5) one can write

$$\upsilon_f = K^{\neq}\omega_f c_A c_{BC}. \qquad (6)$$

On the other hand, the reaction rate coefficient $k$ is defined by $\upsilon_f = k\, c_A c_{BC}$, and then one finds

$$k = K^{\neq} \omega_f. \tag{7}$$

Thus, the reaction rate coefficient $k$ can be obtained by the equilibrium constant multiplied with the reaction coordinate frequency. Determination of the equilibrium constant depends on the statistical distribution of a system at a stationary state. TST presumes that a thermodynamic equilibrium is maintained between transition state and reactants. Under this assumption, the statistical equilibrium distribution function is a BG distribution, and therefore the equilibrium constant $K^{\neq}$ has the form of an exponential law [1],

$$K^{\neq} = \exp\left(-\beta \sum_i \chi_i \mu_i\right), \tag{8}$$

where $\chi_i$ is the stoichiometric number of $i$th composition, and $\mu_i$ is the chemical potential of $i$th composition, and $\sum_i \chi_i \mu_i$ is the reaction Gibbs free energy. If the reaction Gibbs free energy is less than zero, the equilibrium constant is greater than 1 and the reaction proceeds forward. But if the reaction Gibbs free energy is greater than zero, the equilibrium constant is less than 1 and the reaction proceeds backward. If the reaction Gibbs free energy is equal to zero, the equilibrium constant is equal to 1 and thus no reaction takes place. Conventionally, the chemical potential is written as

$$\mu_i = \varepsilon_i^0 - k_B T \ln Z_i, \tag{9}$$

where $Z_i$ is a partition function of $i$th composition. The first term $\varepsilon_i^0$ on the right side of the equation is a basic free energy determined by the molecular structure, and the second term is the free energy determined by the entropy that is contributed by the probability [23]. The present form of the partition function with a logarithm function is as a result of BG entropy contributed by BG distribution with the exponential form. For the elementary reaction (4), the basic free energies of compositions A, BC and $X^{\neq}$ can be denoted by $\varepsilon_A^0$, $\varepsilon_{BC}^0$ and $\varepsilon_{\neq}^0$, and the corresponding partition functions can be written as $Z_A$, $Z_{BC}$ and $Z^{\neq}$, respectively.

Generally speaking, a chemical reaction is not in a thermodynamic equilibrium but in a nonequilibrium state. In reaction rate theory, what we are interested in is the processes of the evolution from one metastable state to another neighboring state, thus the thermodynamic equilibrium assumption in TST is quite farfetched. In particular, when the system far away from equilibrium does not relax to a thermodynamic equilibrium with a BG distribution, but asymptotically approaches to a stationary nonequilibrium with a power-law distribution, the equilibrium constant expression

Eq.(8) becomes invalid and so the TST reaction rate coefficient Eq.(2) must be modified.

In nonextensive statistical mechanics, the power-law distribution Eq.(3) can be derived using the extremization of Tsallis entropy [5]. In stochastic dynamical theory on Brownian motion in a complex system, Eq.(3) can also be obtained by solving Fokker-Planck equations [4, 24]. When one generalizes BG statistics to nonextensive statistics, the usual exponential and logarithm can be replaced respectively by the $q$-exponential and the $q$-logarithm [5]. Here the $\nu$-exponential [25, 26] can be defined as

$$\exp_\nu x = [1+(\nu-1)x]^{1/(\nu-1)}, \quad (\exp_1 x = e^x), \tag{10}$$

if $1+(\nu-1)x >0$ and as $\exp_\nu x =0$ otherwise. And the inverse function, the $\nu$-logarithm can be defined as

$$\ln_\nu x = \frac{x^{\nu-1}-1}{\nu-1}, \quad (x>0, \ln_1 = \ln x). \tag{11}$$

In this framework, the equilibrium constant Eq. (8) can be generalized by

$$K_\nu^{\neq} = \exp_\nu\left(-\beta \sum_i \chi_i \mu_{\nu,i}\right) = \left[1-(\nu-1)\beta \sum_i \chi_i \mu_{\nu,i}\right]^{1/(\nu-1)}. \tag{12}$$

And correspondingly, the relation, Eq.(9), between the chemical potential and the partition function can be generalized by

$$\mu_{\nu,i} = \varepsilon_i^0 - k_B T \ln_\nu Z_i = \varepsilon_i^0 - k_B T \frac{Z_i^{\nu-1}-1}{\nu-1}, \tag{13}$$

In the limit $\nu \to 1$ Eqs.(12) and (13) are reduced to the standard forms in conventional thermodynamic statistics. Here, it would be helpful to introduce the physical meaning of the power-law parameter $\nu \neq 1$. In 2004, an equation of the parameter $\nu \neq 1$ was found both in the self-gravitating system and in the plasma system, and hence a clear physical explanation for $\nu \neq 1$ was presented firstly [6, 7]. The equation can be written as $k_B \nabla T(r) = -(\nu-1)m\nabla \varphi_g(r)$ for the self-gravitating system [6] and $k_B \nabla T(r) = (\nu-1)e\nabla \varphi_C(r)$ for the plasma system [7], where $T(r)$ is space-dependent temperature, $m$ is mass of particle, $e$ is charge of electron, $\varphi_g(r)$ is a gravitational potential function and $\varphi_C(r)$ is a Coulombian potential function. The equation shows that the $\nu$-parameter is $\nu \neq 1$ if and only if $\nabla T(r) \neq 0$ and hereby the power-law distribution represents the nature of an interacting many-body system being at a nonequilibrium stationary-state. For a chemical reaction system, the equation of the $\nu$-parameter should be similar to that for the self-gravitating system (although one needs to study the precise expression), and $\varphi_g(r)$ should be construed as an intermolecular interaction potential function.

Further, substituting Eq. (13) into Eq. (12), one finds

$$K_\nu^{\ne} = \left[1-(\nu-1)\sum_i \chi_i\left(\frac{\varepsilon_i^0}{k_B T} - \frac{Z_i^{\nu-1}-1}{\nu-1}\right)\right]^{1/(\nu-1)}$$

$$= \left[1-(\nu-1)\left(\frac{Z_A^{\nu-1}-1}{\nu-1} + \frac{Z_{BC}^{\nu-1}-1}{\nu-1} - \frac{(Z^{\ne})^{\nu-1}-1}{\nu-1} + \frac{\Delta\varepsilon_0}{k_B T}\right)\right]^{1/(\nu-1)}, \quad (14)$$

where $\Delta\varepsilon_0 = \varepsilon_{\ne}^0 - \varepsilon_A^0 - \varepsilon_{BC}^0$ is the basic energy difference between activated complex and reactants. Because the activated complex is near the maximum of the potential energy along the reaction coordinate, and any change in the inter-atomic distances leads to a decrease in the potential energy and thereby decomposes the activated complex, this is an unstable relaxation vibration. Generally, this kind of motion can be treated as motion of a one-dimensional harmonic oscillator, whose partition function can be replaced [1] by

$$Z_\omega^{\ne} = \left[1-\exp_\nu(-h\omega_f/k_B T)\right]^{-1} \approx \frac{k_B T}{h\omega_f}, \quad (15)$$

where the vibration frequency $\omega_f$ is the same as the reaction coordinate frequency in Eq. (6). The reaction coordinate motion can be separated from the whole motion in the transition state, therefore the whole partition function $Z^{\ne}$ for the transition state can be written as $Z^{\ne} = Z_\omega^{\ne} Z_\Delta^{\ne}$, where $Z_\Delta^{\ne}$ is the transition state partition function which removed the vibration partition function $Z_\omega^{\ne}$. The above factorization of the partition functions also holds for the power-law distributions in nonextensive statistics [19, 25].

Combining Eq.(14) and Eq.(15) with Eq.(7), we can find the generalized TST reaction rate formula for the elementary reaction taking place in the system with the power-law $\nu$-distribution,

$$k_\nu = \omega_f\left[2 + \left(\frac{k_B T}{h\omega_f}\right)^{\nu-1}(Z_\Delta^{\ne})^{\nu-1} - Z_A^{\nu-1} - Z_{BC}^{\nu-1} - (\nu-1)\frac{\Delta\varepsilon_0}{k_B T}\right]^{\frac{1}{\nu-1}}. \quad (16)$$

As expected, in the limit $\nu \to 1$ Eq.(16) is reduced to the standard TST reaction rate formula for the elementary reaction in systems with a BG distribution [1, 3],

$$k_1 = k_{TST} = \frac{k_B T}{h}\frac{Z_\Delta^{\ne}}{Z_A Z_{BC}}\exp\left(-\frac{\Delta\varepsilon_0}{k_B T}\right). \quad (17)$$

We find that the power-law TST reaction rate formula Eq.(16) not only depends on the $\nu$-parameter, but also on the reaction coordinate frequency $\omega_f$ which does not appear in the conventional TST reaction rate formula Eq.(17).

## 3. Numerical analyses of the power-law TST rate coefficient

In order to illustrate dependence of the power-law TST rate coefficient on the physical quantities such as the power-law parameter $\nu \neq 1$, the reaction coordinate frequency $\omega_f$, the temperature $T$, and the activated energy $\Delta\varepsilon_0$, we have made numerical analyses of $k_\nu$ with regard to $\nu$, $\omega_f$, $T$, and $\Delta\varepsilon_0$, respectively, and have studied the variation of the generalized TST rate coefficient $k_\nu$ in Eq.(16) as a function of these quantities. In these numerical analyses, when one of these quantities was chosen as a variable, the other quantities were fixed. The fixed data were taken from the reaction, F + H$_2$ → HF + H, which will be studied as an example of the application in Section 4. In this way, we have chosen $\omega_f$=723 cm$^{-1}$ and $\Delta\varepsilon_0 = 0.7572 \times 10^{-20}$J as the fixed values of the reaction coordinate frequency and the activated energy, and have chosen at the temperature $T$=300K $Z_\Delta^{\neq} = 4.69 \times 10^{34}$, $Z_A = Z_F = 3.47 \times 10^{32}$ and $Z_{BC} = Z_{H_2} = 4.79 \times 10^{30}$ as the fixed values of the partition functions [27]. Except in Fig.3, all the numerical analyses in the figures were carried out at a temperature of $T$=300K.

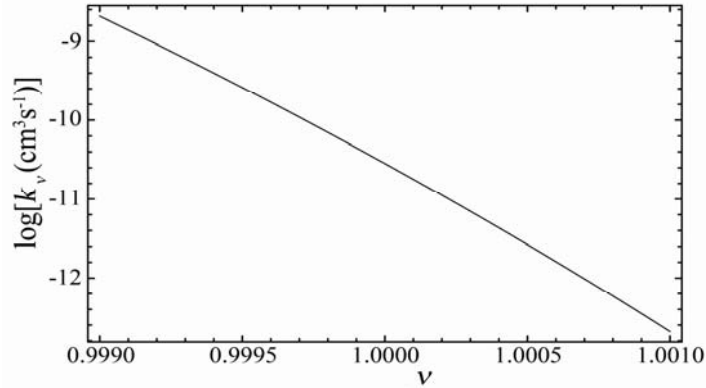

Fig. 1. Dependence of the rate coefficient $k_\nu$ on the parameter $\nu$

Fig.1 has shown dependence of the rate coefficient $k_\nu$ on the parameter $\nu$. The $k_\nu$-axis was plotted on a logarithmic scale. The range of the $\nu$-axis was chosen near 1.000, implying a state not very far away from the equilibrium. Inside the figure, we laid out a magnified plot in the range of the $\nu$-axis: 0.99990~1.00010, a very small deviation from $\nu$=1. In Fig.1, the numerical analyses showed a very strong dependence of the generalized TST rate coefficient $k_\nu$ on the parameter $\nu$, which imply that a tiny deviation from the BG distribution and thus from the thermal equilibrium would result in a significant variation in the reaction rate. Such high sensitivity of the reaction rate to the $\nu$-parameter has shown the important role of the power-law distribution in the calculation of reaction rate coefficient, and again has told us that the nonequilibrium or the statistical distribution function is a key factor considered in the construction of the reaction rate theory.

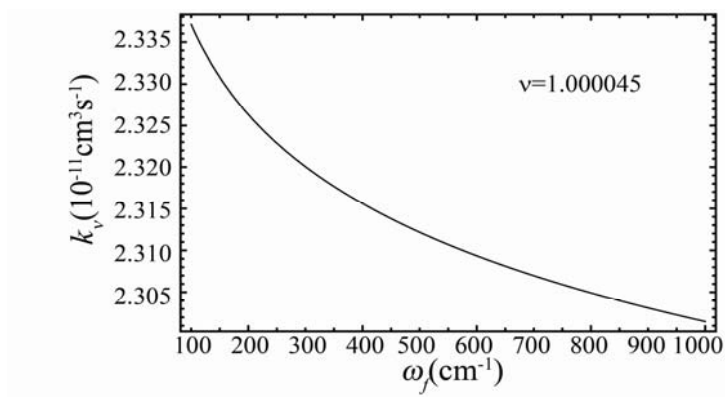

(a) The case $\nu >1$

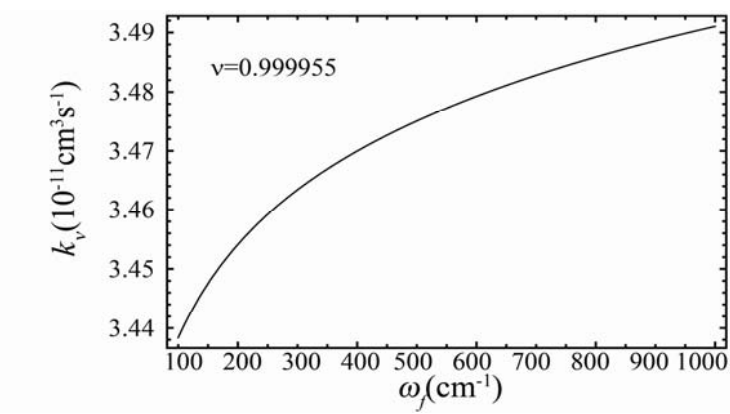

(b) The case $\nu <1$

Fig. 2. (a) and (b). Dependence of rate coefficient $k_\nu$ on reaction coordinate frequency $\omega_f$

In Fig.2 (a) and (b), we illustrated the dependence of the generalized TST rate coefficient $k_\nu$ on the reaction coordinate frequency $\omega_f$ of the activated complex, where two fixed values of the ν-parameter were chosen, respectively, ν=1.000045 in (a) for the case ν >1 and ν=0.999955 in (b) for the case ν <1. The range of the $\omega_f$-axis was chosen 100~1000$cm^{-1}$, being in the order of the fixed value $\omega_f$=723 $cm^{-1}$. It is shown that the dependences of $k_\nu$ on $\omega_f$ for the case ν >1 and for the case ν < 1 are different evidently from each other. For the case ν >1 in (a), the reaction rate decreases gradually as $\omega_f$ increases, but for the case ν < 1 in (b), the reaction rate increases gradually as $\omega_f$ increases. The reason is because the term, $\left(k_B T/h\omega_f\right)^{\nu-1}$, appearing inside the rate formula Eq.(16) has different characteristics for these two cases ν >1 and ν < 1.

The role of the reaction coordinate frequency $\omega_f$ in the reaction rate is worth noticing because it does not appear in the conventional TST rate formula (i.e. the case

ν=1). The reaction coordinate frequency and the potential energy surface can be calculated by using *ab initio* methods (first principles), such as the three dimensional potential energy surface for F+$H_2$→HF+H reaction and the *ab initio* Stark-Werner potential energy surface [27-29].

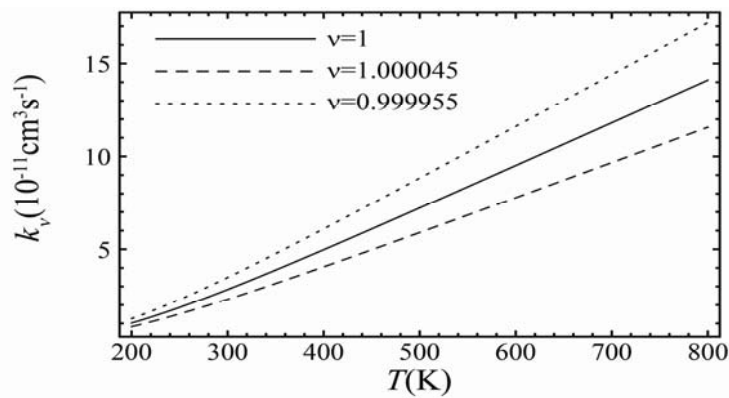

Fig. 3. Dependence of the rate coefficient $k_\nu$ on temperature $T$ for three values of $\nu$

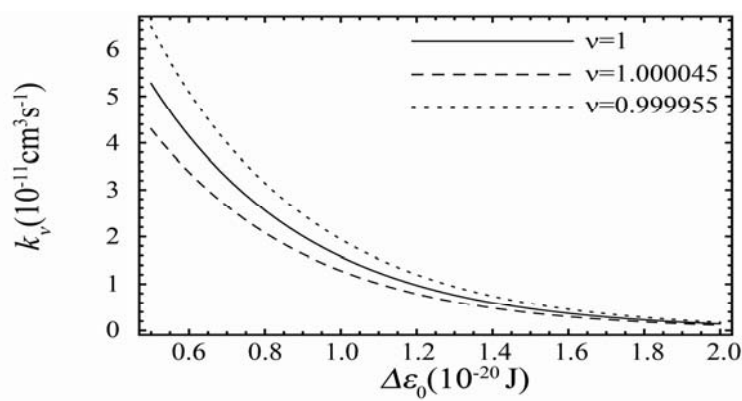

Fig. 4. Dependence of the rate coefficient $k_\nu$ on $\Delta\varepsilon_0$ for three values of $\nu$

Fig.3 illustrated the dependence of the generalized TST rate coefficient $k_\nu$ on temperature $T$ for three values of $\nu$. The range of $T$-axis was chosen 200~800K, extending the temperature range of all the experimental studies in the NIST chemical kinetics database at http://kinetics.nist.gov/kinetics.

Fig.4 illustrated the dependence of the generalized TST rate coefficient $k_\nu$ on the activation energy $\Delta\varepsilon_0$ for three values of $\nu$. The range of $\Delta\varepsilon_0$-axis was chosen 0.5~2 ×$10^{-20}$J, being in the order of the fixed value $\Delta\varepsilon_0 = 0.7572 \times 10^{-20}$J.

In the Figs.(3) and (4), the curves of $\nu$=1 are corresponding to the conventional TST reaction rate coefficient in BG statistics.

## 4. Application to F+$H_2$→HF+H reaction

In order to illustrate the application of the generalized TST reaction rate formula Eq.(16) to the elementary bimolecular reaction occurring in a nonequilibrium system with the power-law ν-distribution, we take the F+$H_2$ reaction as an example to calculate the power-law TST reaction rate coefficient. The elementary reaction process can be written [27] as

$$F+H_2 \xrightarrow{k_v} HF+H. \qquad (18)$$

Table 1. Properties of reactants and activated complex for F+H$_2$ reaction [27-29]

| Parameters | F$\cdots$H$\cdots$H | F | H$_2$ |
|---|---|---|---|
| $r$(F-H)/nm | 0.1546 | | |
| $r$(H-H)/nm | 0.0771 | | 0.07417 |
| $\omega_1$/cm$^{-1}$ | 3772 | | 4395.2 |
| $\omega_2$/cm$^{-1}$ | 296 | | |
| $\omega_3$/cm$^{-1}$ | 723$i$ | | |
| $\Delta\varepsilon_0$/10$^{-20}$J | 0.7572 | | |
| $m$/10$^{-26}$kg | 3.4895 | 3.1548 | 0.3348 |
| $I$/(10$^{-48}$ kg$\cdot$m$^2$) | | | 4.600 |
| $I_a$/(10$^{-48}$ kg$\cdot$m$^2$) | 2.873 | | |
| $I_b$/(10$^{-48}$ kg$\cdot$m$^2$) | 96.43 | | |
| $I_c$/(10$^{-48}$ kg$\cdot$m$^2$) | 99.32 | | |
| $g_{e,0}$ | 2 | 4 | 1 |
| $\sigma$ | 1 | | 2 |

The activated complex (i.e., the transition state) is F$\cdots$H$\cdots$H. Using the London-Eyring-Polanyi-Sato (LEPS) method, the properties of the reactants and the activated complex were listed in Table 1. In this table, there are two real vibration frequencies: $\omega_1$ is the H-H stretching vibration relative to F, and $\omega_2$ is the bending vibration. There is also a virtual vibration frequency of the reaction coordinate, $\omega_3$, marked with $i$ in the table. $I$, $I_a$, $I_b$, and $I_c$ are the moments of inertia, $g_{e,0}$ is the ground state degeneracy (the degeneracy of F at the ground state $^2P_{3/2}$ is 4, the degeneracy of F$\cdots$H$\cdots$H at the ground state is 2, and the degeneracy of H$_2$ at the ground state $^1\Sigma_g^+$ is 1). The symmetry number is $\sigma$ for calculations of the rotations.

The molecular movements are composed of five parts: the translation, the rotation, the vibration, and the motion of electrons and nucleus. Usually, the motion of nucleus can be ignored because a general chemical reaction does not change nuclear structure. In addition, assuming the motions that are localized inside a molecule follow BG statistics and assuming separability of the partition function, one can use molecular partition functions of the conventional forms in BG statistics. Therefore, they can generally be written [1] as $Z=Z_tZ_rZ_vZ_e$, where the translation partition function is

$$Z_t = \left(\frac{2\pi m k_B T}{h^2}\right)^{3/2}, \tag{19}$$

the rotation partition function is

$$Z_r = \begin{cases} \dfrac{8\pi^2 I k_B T}{\sigma h^2}, & \text{for diatomic or linear polyatomic molecule,} \\ \dfrac{\sqrt{\pi}}{\sigma}\left(\dfrac{8\pi^2 k_B T}{h^2}\right)^{3/2}\sqrt{I_a I_b I_c}, & \text{for nonlinear polyatomic molecule,} \end{cases} \tag{20}$$

the vibration partition function is

$$Z_v = \prod_{i=1}^{s}\left[1-\exp\left(-\frac{\hbar\omega_i}{k_B T}\right)\right]^{-1} \tag{21}$$

with $s=3n-5$ for the linear molecule and $s=3n-6$ for the nonlinear molecule, and the electron partition function is

$$Z_e = g_{e,0} + g_{e,1} e^{-(E_{e,1}-E_{e,0})/k_B T} + \cdots. \tag{22}$$

Now back to the application of formula Eq. (16) to the reaction process (18), one has $Z_A = Z_F$, $Z_{BC} = Z_{H_2}$, and $Z_\Delta^{\neq} = Z_{F\cdots H\cdots H}$, and then obtains

$$Z_A = Z_{At}Z_{Ae} = \left(\frac{2\pi m_A k_B T}{h^2}\right)^{3/2}\left[4+2\exp\left(-\Delta E/k_B T\right)\right] = 3.47\times 10^{32}, \tag{23}$$

$$Z_{BC} = Z_{BCt}Z_{BCr}Z_{BCv}Z_{BCe}$$
$$= \left(\frac{2\pi m_{BC} k_B T}{h^2}\right)^{3/2}\frac{8\pi^2 I_{BC} k_B T}{\sigma_{BC} h^2}\left[1-\exp\left(-\frac{\hbar\omega_1}{k_B T}\right)\right]^{-1} g_{e0} = 4.79\times 10^{30}, \tag{24}$$

and $Z_\Delta^{\neq} = Z_{\neq t}Z_{\neq r}Z_{\neq v}Z_{\neq e}$

$$= \left(\frac{2\pi m_{\neq} k_B T}{h^2}\right)^{3/2}\frac{\sqrt{\pi}}{\sigma_{\neq}}\left(\frac{8\pi^2 k_B T}{h^2}\right)^{3/2}\sqrt{I_a I_b I_c}\prod_{i=1}^{2}\left[1-\exp\left(-\frac{\hbar\omega_i}{k_B T}\right)\right]^{-1} g_{e0} = 4.69\times 10^{34}, \tag{25}$$

where in Eq.(23). It is worth mentioning that the spin–orbit coupling has been taken into account, calculated by the approximate equation, $4 + 2\exp(-\Delta E/k_B T)$. The energy difference $\Delta E$ between the first excited state and the ground state is taken 0.05eV [27].

The rate coefficients of reaction (18) in experimental measurements were taken from NIST chemical kinetics database at http://kinetics.nist.gov/kinetics. All of the experimental studies were carried out in temperature range: 190 ~ 765K, and the reaction rate coefficients in the experimental measurements were satisfied by the two-parameter fit, namely

$$k_{\exp} = 1.29\times 10^{-10} cm^3 s^{-1} \exp(-517.17\,K/T) \tag{26}$$

with RMSD=0.5. In Table 2, we listed the experimental values and the theoretical values in the temperature range 190~765K of the rate coefficient of the reaction (18), where $k_{TST}$ was calculated using Eq.(17), $k_{exp}$ was obtained using Eq.(26), and $k_\nu$ was calculated using Eq.(16). The quantity $\delta$ was used to denote the relative error of $k_{TST}$ to $k_{exp}$, i.e. $\delta=|k_{TST} - k_{exp}| / k_{exp}$. The composition frequency $\omega_f$ is $\omega_3$ in Table 1. The results were listed in Table 2.

Table 2. The experimental values and the theoretical values of
the rate coefficients of $F+H_2$ reaction

| $T$(K) | $k_{TST}$($10^{-11}$cm$^3$s$^{-1}$) | $k_{exp}$($10^{-11}$cm$^3$s$^{-1}$) | $\delta$ | $k_\nu$($10^{-11}$cm$^3$s$^{-1}$) | $\nu$ |
|---|---|---|---|---|---|
| 190 | 0.88 | 0.85 | 3.5% | 0.85 | 1.000006 |
| 300 | 2.84 | 2.30 | 23% | 2.30 | 1.000045 |
| 400 | 4.98 | 3.54 | 41% | 3.54 | 1.000075 |
| 500 | 7.22 | 4.59 | 57% | 4.59 | 1.000101 |
| 600 | 9.51 | 5.45 | 74% | 5.45 | 1.000124 |
| 700 | 11.8 | 6.16 | 92% | 6.16 | 1.000145 |
| 765 | 13.3 | 6.56 | 103% | 6.56 | 1.000158 |

Table 2 shows significant relative errors of $k_{TST}$ to $k_{exp}$, but the values of $k_\nu$ with different $\nu$-parameter which are exactly agreement with all the experimental studies in the temperature range.

We find that there is one ν-parameter in the experimental measurements at one temperature, and the ν-parameter varies as the temperature varies. Such a variation in the ν-parameter is a result of the fact that the ν-parameter may not only depend on the intermolecular interactions but also on the temperature, which mirrors the differences between the experiments at different temperatures and the environment. In fact, by studying the Brownian motion in a complex medium one has predicted that the power-law parameter measures a distance away from thermal equilibrium [4]. According to Du's equation [6, 7], the power-law parameter different from unity is related to temperature gradient and interaction potential gradient in a nonequilibrium system.

Although the tunneling effect for the chemical reaction was not taken into account in present calculations, it would not have any effect on the validity of the generalized TST reaction rate formula Eq.(16). If the tunneling effect is considered in the rate calculation for this reaction, it will result in a slightly different calculation value of the ν-parameter.

## 5. Conclusion and discussion

The reaction rate theory for the systems with power-law distributions is beyond the scope of conventional TST for the systems with a BG distribution, and therefore if chemical reactions occur in the systems with power-law distributions the TST reaction rate formulae need to be modified. In conclusion, we have studied the TST reaction rate coefficient of an elementary reaction taking place in a nonequilibrium system with the power-law $\nu$-distribution. We have derived a generalized TST reaction rate formula Eq.(16), which, as compared with the old formula Eq.(17), Eq.(16) not only depends on the power-law parameter $\nu$, but also on the reaction coordinate frequency $\omega_f$.

The purpose of this work is to generalize the conventional TST rate formula Eq.(17) or Eq.(2) to a nonequilibrium system with the power-law ν-distribution. The present work is only a classical statistical theory, and the approach to generalize the TST rate formula follows the standard line in general textbooks. In this way, the new rate coefficient Eq.(16) and the old one Eq.(17) are parallel to the factors introduced by the quantum effects such as the recrossing effect and the tunneling effect. In other words, if the recrossing coefficient and the tunneling coefficient were incorporated into Eq.(17), they should also be incorporated into Eq.(16). For example, if the old rate coefficient was written as $\gamma \kappa k_{TST}$ with a recrossing coefficient $\gamma$ and a tunneling coefficient $\kappa$, then the new rate coefficient could be written as $\gamma \kappa k_\nu$, which does not affect the validity of the power-law TST rate formula Eq.(16).

We have made numerical analyses to illustrate the dependence of the power-law TST reaction rate coefficient $k_\nu$ on the relevant physical quantities. We clearly showed a very strong dependence of $k_\nu$ on the power-law $\nu$-parameter, and indicated that a tiny deviation from the BG distribution (thus from thermodynamic equilibrium) would result in a very significant effect on the reaction rate. Such high sensitivity of the reaction rate coefficient to the parameter ν showed the important role of the power-law distributions in the calculations of reaction rate coefficients, and again told us that the nonequilibrium is a key factor to be considered in the construction of the reaction rate theory for a complex system. In addition to the usual quantities, such as the temperature $T$ and the activated energy $\Delta \varepsilon_0$, contained in the rate coefficient, In particular we have analyzed the effect of $\omega_f$ on $k_\nu$, and in Fig.2 we have shown an important difference of this effect between the case of ν>1 and the case of ν<1.

In order to illustrate the application of the power-law TST reaction rate formula Eq.(16) to an elementary reaction taking place in a nonequilibrium system with the power-law ν-distribution, we have taken the reaction of fluorine and hydrogen, F+$H_2$ →FH+H , as an example to calculate the power-law TST reaction rate coefficient $k_\nu$. With different values of the $\nu$-parameter, the calculation values of the rate coefficient $k_\nu$ for this reaction are exactly agreement with the measurements in all the experimental studies in the temperature range of 190~765K.

*Additional remark*: there has been a great deal of work done to develop models of phenomenological kinetics with non-BG distributions. In the field of gas phase kinetics, much of the work has been aimed at formulating energy grained kinetic master equations [30~33], showing that small perturbations in BG distribution can have significant effects on the rate coefficients.

**Acknowledgment**

This work is supported by the National Natural Science Foundation of China under Grant No. 11175128 and also by the Higher School Specialized Research Fund for Doctoral Program under Grant No. 20110032110058.